\documentclass[letter,prb,superscriptaddress,showpacs,floatfix,twocolumn]{revtex4}

\usepackage[dvips]{graphics} 
\usepackage{graphicx}

\begin{document}

\title{The Korringa-Kohn-Rostoker Non-Local Coherent Potential Approximation \\ (KKR-NLCPA)}

\author{D.A.Rowlands}
\email{d.a.rowlands@warwick.ac.uk}
\affiliation{Dept. of Physics, University of Warwick, Coventry CV4 7AL, U.K.}
\author{J.B.Staunton}
\affiliation{Dept. of Physics, University of Warwick, Coventry CV4 7AL, U.K.}
\author{B.L.Gy\"{o}rffy}
\affiliation{H.H.Wills Physics Laboratory, University of Bristol, Bristol BS8 1TL, U.K.}
   
\date{\today}   
       

\begin{abstract}
We introduce the Korringa-Kohn-Rostocker non-local coherent potential approximation (KKR-NLCPA) for describing the electronic structure of
disordered systems. The KKR-NLCPA systematically provides a hierarchy of improvements upon the widely used KKR-CPA approach and includes
non-local correlations in the disorder configurations by means of a self-consistently embedded cluster. The KKR-NLCPA method satisfies all 
of the requirements for a successful cluster generalization of the KKR-CPA; it remains fully causal, becomes exact in the limit of large 
cluster sizes, reduces to the KKR-CPA for a single-site cluster, is straightforward to implement numerically, and enables the effects of
short-range order upon the electronic structure to be investigated. In particular, it is suitable for combination with electronic density
functional theory to give an ab-initio description of disordered systems. Future applications to charge correlation and lattice
displacement effects in alloys and spin fluctuations in magnets amongst others are very promising. We illustrate the method by application
to a simple one-dimensional model. 

\end{abstract}


\maketitle


\section{Introduction}

Over the past 30 years or so the coherent potential approximation~\cite{Soven} (CPA) has proved to be a generally reliable method for dealing 
with disordered systems.~\cite{Elliott,Eh:Sc} However, being in essence a single-site mean-field theory,~\cite{Va:Vo} the CPA fails to take 
into account the effect of non-local potential correlations due to the disorder in the environment of each site and hence leaves much 
important physics out of consideration. Consequently, considerable effort has been spent in trying to find a way of improving it 
systematically by a multi-site or cluster generalization. Surprisingly this has turned out to a very difficult problem~\cite{Gonis} and a
viable solution has been proposed only recently. The new method has emerged from the Dynamical Cluster 
Approximation~\cite{Ja:Rapid,Ja:DCA,Maier} (DCA) which was directed originally at describing dynamical spin and charge fluctuations in simple
Hubbard models of strongly-correlated electron systems. Recently its static limit has been adapted by Jarrell and Krishnamurthy for a simple 
tight-binding model of electrons moving in a disordered potential.~\cite{Ja:Kr} The same problem was investigated by Moradian et
al.~\cite{Mor} In this paper we develop the ideas behind this approximation further and demonstrate how they can be combined with realistic,
ab-initio descriptions of systems of interacting electrons in disordered systems.

Because the language of our multiple scattering theory is so different from that of the context in which the DCA is usually deployed, we 
elaborate on this relationship. Firstly, we note that the DCA was invented to describe short-range correlations within the framework of the
Dynamical Mean Field Theory~\cite{Georges} (DMFT) of spin and charge fluctuations in many-electron systems. Secondly, we recall that that the
DMFT can be regarded as the dynamical generalization of the CPA for the Hubbard `alloy analogy' problem.~\cite{Hub,Kake} Thus, in the light 
of these remarks, it is natural to investigate the static version of the DCA as a generalization of the CPA which includes a description of 
short-range order. Indeed, Jarrell and Krishnamurthy~\cite{Ja:Kr} already studied the problem of electronic structure in random alloys from 
this point of view. In this paper we tackle the same problem using an identical conceptual framework but a very different description of the 
electrons afforded by multiple scattering theory.~\cite{Gon:But,Wein} To make the above remarks more specific we would like to highlight two
of the principal differences between our treatment of the problem and that of Ref.~\onlinecite{Ja:Kr}. Firstly, we do not make use of a 
tight-binding model Hamiltonian but solve, numerically, a Schr\"{o}dinger equation in each unit cell and match the `out-going wave' solution
to the incoming waves from all the other unit cells. This is known as the multiple scattering approach~\cite{Gon:But,Wein} to the problem of 
electronic structure in solids and is the foundation of the Korringa-Kohn and Rostoker (KKR) band theory method. Consequently, the principal
virtue of our formalism, as opposed to those based on tight-binding model Hamiltonians is that it prepares the ground for first-principles 
calculations based on density-functional theories.~\cite{Dre:Gro} The second difference is a formal consequence of the first. In multiple 
scattering theories the object of interest is not the self-energy and the diagrammatic language of perturbation theory is not used. We will 
show that the quantities that play the role of the self-energy in multiple scattering theory are the effective scattering amplitudes 
\,$\widehat{\underline{t}}$\, and effective structure constants \,$\widehat{\underline{G}}(\mathbf{R}^{ij})$\, which are also
the natural concepts in effective medium theories.~\cite{Choy} In short, these formal reasons fully account for the fact that we do not base
our arguments on `restoring momentum conservation' and introducing approximate Laue functions to renormalize diagrams but construct our 
theory in terms of real and reciprocal space clusters. Nevertheless, we believe that our final algorithm described in Sec.~\ref{algorithm}
is equivalent to those investigated by Jarrell and Krishnamurthy.~\cite{Ja:Kr} Our aim in reformulating the problem is to facilitate the 
deployment of the method as a first-principles calculation, in other words to develop a non-local KKR-CPA.~\cite{Gyorffy}   

In brief, our KKR-NLCPA method introduces new effective structure constants and this enables us to define an effective medium which includes 
non-local potential correlations over all length scales. Using a `coarse-graining' procedure inspired by the DCA we can then derive
a self-consistent `cluster generalization' of the KKR-CPA~\cite{Gyorffy} which determines an approximation to this effective medium by 
including non-local correlations up to the range of the cluster size. The KKR-NLCPA satisfies all of the requirements for a successful 
cluster generalization of the KKR-CPA as listed by Gonis.~\cite{Gonis} In particular, the KKR-NLCPA becomes exact in the limit of large 
cluster sizes where it includes non-local correlations over all length scales, and recovers the KKR-CPA for a single-site cluster. The
method is fully causal, allows the effects of short-range order to be modelled, and can be implemented numerically for realistic systems.

The outline of this paper is as follows. In the next section we describe the formalism for the KKR-NLCPA. We explain our KKR-NLCPA algorithm
and show how to include short-range order. We describe in more detail how to carry out the coarse-graining with reference to simple cubic,
body-centered cubic and face-centered cubic lattices. Finally we explain how to use the KKR-NLCPA formalism to calculate observable 
quantities such as the configurationally-averaged density of states in preparation for DFT calculations. In order to illustrate the 
improvements over the conventional KKR-CPA, in Sec.~\ref{results} we present results (configurationally-averaged density of states) for the 
application of the formalism to a one-dimensional model. However we emphasise that the formalism presented is fully tractable for realistic 
three-dimensional systems.


\section{Formalism}


\subsection{The KKR-CPA} 

For the sake of clarity, we begin by briefly summarising the idea of the conventional KKR-CPA~\cite{Gyorffy,Nato} method. The path operator
equation describing the scattering of an electron in a general array of non-overlapping potentials, centered on site positions 
$\{\mathbf{R}_i\}$, is given by
\begin{equation} 
	\underline{\tau}\,^{ij}=\underline{t}\,^i\delta_{ij}+\sum_{k\neq i}
	\underline{t}\,^{i}\,\underline{G}(\mathbf{R}^{ik})\,\underline{\tau}\,^{kj}    
\end{equation}
where the underscore denotes a matrix in angular momentum space with the usual angular momentum indices \,$l,m$\, and 
\,$\mathbf{R}^{ij}=\mathbf{R}_i-\mathbf{R}_j$\,
is the position vector connecting sites $i$ and $j$. For spherically symmetric scatterers the individual t-matrices are diagonal i.e.
\,$\underline{t}^i=t^i_{lm}\delta_{ij}\delta_{lm,l'm'}$\, but the structure constants \,$G_{lm,l'm'}(\mathbf{R}^{ik})$\, are not.
We shall be interested in situations where the lattice sites labeled by $i$ and $j$ form a regular array, an infinite lattice, but the 
single-site t-matrices vary from site to site in a random fashion. If we consider the solution of Eq.~(1) for all possible disorder 
configurations and then take its average, we arrive at an averaged path operator \,$\overline{\underline{\tau}}\,^{ij}$\, which is 
translationally-invariant. In practice it is not computationally feasible to average over every possible disorder configuration nor is an 
equation which determines the averaged path operator \,$\overline{\underline{\tau}}\,^{ij}$\, readily available in the form of Eq.~1. 
Under these circumstances a way forward is to follow the strategy of `effective medium theories'.~\cite{Ziman,Choy} In the present context 
a useful `effective medium' is that provided by an ordered array of effective scatterers described by the same t-matrix
\,$\widehat{\underline{t}}$\,. In the KKR-CPA,~\cite{Gyorffy,Nato} the scattering amplitude describing these effective scatterers is 
determined using the self-consistency condition that excess scattering off a single-site impurity embedded in such a medium be zero on the 
average. As mentioned in the introduction, whilst very successful in many applications, being a single-site approximation the CPA fails to
take into account the effect of non-local potential correlations due to the disorder in the environment of each site. 


\subsection{Inclusion of non-local potential correlations}

The first step in going beyond the KKR-CPA is to define what we will call the non-local CPA (NLCPA) effective medium by the following
equation:
\begin{equation}
	\widehat{\underline{\tau}}\,^{ij}=\widehat{\underline{t}}\,\delta_{ij}+\sum_{k\neq i}
	\widehat{\underline{t}}\,\widehat{\underline{G}}(\mathbf{R}^{ik})\,\widehat{\underline{\tau}}\,^{kj} 
\end{equation}
where a circumflex symbol denotes an NLCPA effective medium quantity. Here we have defined NLCPA effective local t-matrices
\,$\widehat{\underline{t}}$\, and a new \emph{effective propagator} by
\begin{equation}
	\widehat{\underline{G}}(\mathbf{R}^{ij})=\underline{G}(\mathbf{R}^{ij})+\widehat{\underline{\alpha}}\,^{ij}
\end{equation}
This is composed of the usual free-space KKR structure constants \,$\underline{G}(\mathbf{R}^{ij})$\, which account for the lattice 
structure plus a \emph{translationally-invariant} effective disorder term 
\,$\widehat{\underline{\alpha}}\,^{ij}\,(\equiv\widehat{\alpha}\,^{ij}_{lm,l'm'})$\,. The matrix 
\,$\widehat{\underline{\alpha}}\,^{ij}$\, takes into account, in an averaged manner, the non-local correlations due to the disorder 
configurations. 

Clearly a cluster generalization of the KKR-CPA would involve determining the above NLCPA effective medium by requiring excess scattering 
off an impurity \emph{cluster} of real potentials embedded in the effective medium to be zero on the average, and indeed this is the
strategy followed in the next section. However the fact that \,$\widehat{\underline{\alpha}}\,^{ij}$\, is a translationally-invariant 
effective medium quantity means that it is diagonal in \,$\textbf{k}$\,, reciprocal space. In the present problem it is important to 
treat the theory in both real lattice and reciprocal lattice space consistently and crucially we shall deal with this by a coarse-graining 
procedure. 


\subsection{Cluster generalization of the KKR-CPA}

The first step is to consider a cluster of sites in the NLCPA effective medium. Denoting the sites within the cluster by capital letters, 
it can be shown (see Appendix A) that Eq.~(2) can be rewritten as a cluster equation given by
\begin{equation}
	\widehat{\underline{\tau}}\,^{IJ}=\widehat{\underline{t}}_{cl}\,^{IJ} +\sum_{K,L}
	\widehat{\underline{t}}_{cl}\,^{IK}\,\widehat{\underline{\Delta}}\,^{KL}\,\widehat{\underline{\tau}}\,^{LJ} 
\end{equation}
with the effective \emph{cluster t-matrix} defined as
\begin{equation}
	\widehat{\underline{t}}_{cl}\,^{IJ}=\widehat{\underline{t}}\,\delta_{IJ}+\sum_{K\neq I}
	\widehat{\underline{t}}\,\widehat{\underline{G}}(\mathbf{R}^{IK})\,\widehat{\underline{t}}_{cl}\,^{KJ}
\end{equation}
Eq.~(4) is simply a re-arrangement of Eq.~(2) so that the site matrix elements of all matrices involve the cluster sites only. 
The cluster t-matrix \,$\widehat{\underline{t}}_{cl}\,^{IJ}$\, describes the scattering within a cluster and
\,$\widehat{\underline{\tau}}\,^{IJ}$\, takes account of all scatterings outside of the cluster via the effective 
\emph{cluster renormalised interactor}~\cite{Gonis,Ziman} $\widehat{\underline{\Delta}}\,^{IJ}$. Nevertheless, since
\,$\widehat{\underline{G}}(\mathbf{R}^{IJ})=\underline{G}(\mathbf{R}^{IJ})+\widehat{\underline{\alpha}}\,^{IJ}$\,, it is clear that the 
cluster t-matrix also includes non-local correlations between the cluster sites.

It should be stressed that \,$\widehat{\underline{\Delta}}\,^{IJ}$\, describes all paths from the cluster site $I$ to the cluster site $J$
which only involve intermediate sites outside of the cluster. It is independent of the contents of the cluster and can be viewed as 
describing the effective medium from which the cluster has been removed i.e. replaced by a \emph{cavity}. We may now define an `impurity 
cluster' embedded in the effective medium simply by filling up this cavity with a particular configuration of site potentials. Clearly, 
the path operator for sites $I,J$ belonging to such an impurity cluster is given by  
\begin{equation}
	\underline{\tau}_{imp}^{\,IJ}=\underline{t}_{cl,imp}^{\,IJ}+\sum_{K,L}
	\underline{t}_{cl,imp}^{\,IK}\, \widehat{\underline{\Delta}}\,^{KL}\, \underline{\tau}_{imp}^{\,LJ}
\end{equation}
with the impurity cluster t-matrix defined by
\begin{equation}
	\underline{t}_{cl,imp}^{\,IJ}=\underline{t}\,^I\delta_{IJ} +\sum_{K\neq I}
	\underline{t}\,^I \,\underline{G}(\mathbf{R}^{IK})\, \underline{t}_{cl,imp}^{\,KJ}
\end{equation}
For a cluster containing $N_c$ sites each scattering according to $\underline{t}^A$ or $\underline{t}^B$, there are $2^{N_c}$ possible 
impurity cluster configurations.

We are now in a position to generalize the usual CPA self-consistency condition to determine our approximation to the exact 
configurationally-averaged medium described by \,$\overline{\underline{\tau}}\,^{IJ}$\,, namely the NLCPA effective medium described by
\,$\widehat{\underline{\tau}}\,^{IJ}$\,. This follows by considering for each configuration the impurity cluster path operator for paths 
starting and ending on the impurity cluster sites and demanding that the average over all configurations be equal to the path operator for
the NLCPA effective medium itself i.e.
\begin{equation}
	\langle\,\underline{\tau}_{imp}^{\,IJ}\,\rangle=\widehat{\underline{\tau}}\,^{IJ}
\end{equation}
The important point to note is that unlike many other self-consistent cluster theories such as the molecular coherent potential
approximation~\cite{Duc,Ma:Ja:2q} (MCPA) and extensions, here \,$\widehat{\underline{\tau}}\,^{IJ}$\, and \,$\widehat{\underline{t}}$\, 
correspond to an effective medium which is invariant under translation from site to site and hence does not yield spurious gaps in the band 
structure which could affect the calculation of transport properties etc.~\cite{Gonis} This is a consequence of our definition of the 
non-local correlation terms \,$\widehat{\underline{\alpha}}^{ij}$\, as translationally-invariant. To preserve this periodicity, during each
self-consistency cycle we must have
\begin{equation}
	\widehat{\underline{\tau}}\,^{IJ}=\frac{1}{\Omega_{BZ}}\int_{\Omega_{BZ}}\mathbf{dk}
	\left(\widehat{\underline{t}}^{-1}-\widehat{\underline{\alpha}}(\mathbf{k})-\underline{G}(\mathbf{k})\right)^{-1}
	e^{i\mathbf{k}(\mathbf{R}_I-\mathbf{R}_J)}
\end{equation}
This can be seen by applying the usual lattice Fourier transform to Eq.~(2) and then considering the cluster sites $I,J$. Clearly, to carry
out the above integration numerically we must have a specific representation of the function \,$\widehat{\underline{\alpha}}(\mathbf{k})$\,.
Following the central idea of the DCA, we proceed by coarse-graining \,$\widehat{\underline{\alpha}}(\mathbf{k})$\, over the Brillouin zone. 
Unlike many other attempts,~\cite{Gonis} this approach has been shown to be fully causal.~\cite{Ja:DCA}


\subsection{Implementation - coarse-graining of the Brillouin zone}

Due to the finite size $N_c$ of the cluster in real-space, the differences in distance between the cluster sites correspond to a set
of $N_c$ cluster momenta $\{\mathbf{K}_n\}$ in reciprocal space according to the relation
\begin{equation}
	\frac{1}{N_c}\sum_{\mathbf{K}_n}e^{i\mathbf{K}_n(\mathbf{R}_I-\mathbf{R}_J)}=\delta_{IJ}
\end{equation}
These $\{\mathbf{K}_n\}$ are at the centers of a set of $N_c$ reciprocal-space patches which coarse-grain the first Brillouin zone of the 
lattice. Earlier it was noted that the effective cluster t-matrix also includes non-local correlation terms 
\,$\widehat{\underline{\alpha}}^{IJ}$\, acting between the cluster sites. We use the above relation to coarse-grain 
\,$\widehat{\underline{\alpha}}(\mathbf{k})$\, as follows:
\begin{equation}
	\widehat{\underline{\alpha}}\,^{IJ}=\frac{1}{N_c}\sum_{\mathbf{K}_n}\widehat{\underline{\alpha}}(\mathbf{K}_n)
	e^{i\mathbf{K}_n(\mathbf{R}_I-\mathbf{R}_J)}
\end{equation}
\begin{equation}
	\widehat{\underline{\alpha}}(\mathbf{K}_n)=\sum_{J\neq I}\widehat{\underline{\alpha}}\,^{IJ}
	e^{-i\mathbf{K}_n(\mathbf{R}_I-\mathbf{R}_J)}
\end{equation}
Since the NLCPA maps the effective lattice problem to that of a self-consistently embedded impurity cluster problem, the fundamental 
assumption we make is that provided we are aiming to reproduce these correlations of finite range in the effective lattice, we may take the
above coarse-grained values \,$\widehat{\underline{\alpha}}(\mathbf{K}_n)$\, to be a good approximation to 
\,$\widehat{\underline{\alpha}}(\mathbf{k})$\, for the effective lattice. 
In short, \,$\widehat{\underline{\alpha}}(\mathbf{k})=\,\widehat{\underline{\alpha}}(\mathbf{K}_n)$\, if $\mathbf{k}$ is in the $n^{th}$
reciprocal-space patch. This is discussed more formally in Appendix B.

The next step is to define the `coarse-grained averaged' reciprocal-space matrix elements for the effective medium path operator by
\begin{equation}
	\widehat{\underline{\tau}}(\mathbf{K}_n)=\frac{N_c}{\Omega_{BZ}}\int_{\Omega_{\mathbf{K}_n}}\mathbf{dk}
	\left(\widehat{\,\underline{t}}\,^{-1}-\widehat{\underline{\alpha}}(\mathbf{K}_n)-\underline{G}(\mathbf{k})\right)^{-1}
\end{equation}
where each integral is over the reciprocal-space patch $\Omega_{\mathbf{K}_n}$ of volume $N_c/\Omega_{BZ}$ surrounding the point
$\mathbf{K}_n$. This is straightforward since each \,$\widehat{\underline{\alpha}}(\mathbf{K}_n)$\, is constant within its coarse-graining 
patch. Defining transformations relating the real-space and coarse-grained reciprocal-space matrix elements by
\begin{equation}
	\widehat{\underline{\tau}}\,^{IJ}=\frac{1}{N_c}\sum_{\mathbf{K}_n}\widehat{\underline{\tau}}(\mathbf{K}_n)
	e^{i\mathbf{K}_n(\mathbf{R}_I-\mathbf{R}_J)}
\end{equation}
\begin{equation}
	\widehat{\underline{\tau}}(\mathbf{K}_n)=\sum_J\widehat{\underline{\tau}}\,^{IJ}e^{-i\mathbf{K}_n(\mathbf{R}_I-\mathbf{R}_J)}
\end{equation} 
means that the real-space effective medium path operator for sites $I,J$ within the cluster is now given by
\begin{widetext}
\begin{equation}
	\widehat{\underline{\tau}}\ ^{IJ}=\frac{1}{\Omega_{BZ}}\sum_{\mathbf{K}_n}\left(\int_{\Omega_{\mathbf{K}_n}}\mathbf{dk}
	\left(\,\widehat{\underline{t}}\,^{-1}-\widehat{\underline{\alpha}}(\mathbf{K}_n)-\underline{G}(\mathbf{k})\right)^{-1}
	e^{i\mathbf{K}_n(\mathbf{R}_I-\mathbf{R}_J)}\right)
\end{equation}
\end{widetext}
which we use to replace Eq.~(9), and we can now iterate to self-consistency until Eq.~(8) and Eq.~(16) are satisfied.

Finally, note that the NLCPA reduces to the CPA for $N_c=1$. In this limit the non-local correlation terms 
\,$\widehat{\underline{\alpha}}^{IJ}$\, vanish and the NLCPA effective t-matrix becomes equal to the usual CPA effective t-matrix. The NLCPA
becomes exact as $N_c\rightarrow\infty$ where $\mathbf{K}_n\rightarrow\mathbf{k}$ and non-local correlations over all length scales are 
treated. For clarity the full KKR-NLCPA algorithm is now summarised below.


\subsection{KKR-NLCPA Algorithm}

\label{algorithm}

All real-space matrices in the algorithm are super-matrices (denoted by double underscores) in cluster-site and angular momentum space.
For a particular energy $E$,
\begin{enumerate}

\item 
Make a guess for the effective cluster t-matrix \,$\widehat{\underline{t}}_{cl}\,^{IJ}$\, for the first iteration. Do this by placing an
average t-matrix (ATA), \,$\overline{\underline{t}}=P(A)\underline{t}^A+P(B)\underline{t}^B$\, (where P(A) is the probability of a site being
occupied by an $A$ atom), on each cluster site, and for the site to site propagation terms in the cluster use the free-space structure 
constants i.e. set \,$\widehat{\underline{\alpha}}\,^{IJ}=0$\,.

\item 
Calculate \,$\widehat{\underline{\alpha}}\,^{IJ}$\, using Eq.~(5) i.e.
\[ \widehat{\underline{\underline{\alpha}}}=\widehat{\underline{\underline{t}}}\,^{-1}-\widehat{\underline{\underline{t}}}_{cl}\,^{-1}
-\underline{\underline{G}} \]
where \,$\widehat{\underline{\underline{t}}}\,^{-1}$\, is the diagonal part of \,$\widehat{\underline{\underline{t}}}_{cl}\,^{-1}$\,.
For the first iteration $\widehat{\underline{\alpha}}\,^{IJ}$ will of course be zero.

\item
Convert the matrix elements \,$\widehat{\underline{\alpha}}\,^{IJ}$\, to coarse-grained reciprocal space using Eq.~(12).

\item
Calculate the coarse-grained matrix elements \,$\widehat{\underline{\tau}}(\mathbf{K}_n)$\, using Eq.~(13) and convert them to real space 
using Eq.~(14).

\item
Calculate \,$\widehat{\underline{\Delta}}\,^{IJ}$\, by solving Eq.~(4) i.e.
\[ \widehat{\underline{\underline{\Delta}}}=\widehat{\underline{\underline{t}}}_{cl}\,^{-1}-
\widehat{\underline{\underline{\tau}}}\,^{-1} \]

\item 
Calculate \,$\underline{{\tau}}_{imp}\,^{IJ}$\, for each impurity cluster configuration using Eq.~(6) and average over all $2^{N_c}$ 
configurations to obtain a new effective path operator at the cluster sites \,$\widehat{\underline{{\tau}}}\,^{IJ}$\,.

\item
Calculate the new cluster t-matrix \,$\widehat{\underline{t}}_{cl}\,^{IJ}$\, by solving Eq.~(4) using 
\,$\widehat{\underline{{\tau}}}\,^{IJ}$\, above and \,$\widehat{\underline{\Delta}}\,^{IJ}$\, from step 5 i.e.
\[ \widehat{\underline{\underline{t}}}_{cl}=\left(\widehat{\underline{\underline{\tau}}}\,^{-1}
+\widehat{\underline{\underline{\Delta}}}\right)^{-1} \]
   
\item
Compare the new cluster t-matrix elements \,$\widehat{\underline{t}}_{cl}\,^{IJ}$\, with those in step 1. If they are not equal to within
the desired accuracy, repeat as necessary steps 2 $\rightarrow$ 8 using the new cluster t-matrix until convergence within the desired 
accuracy is achieved.
     
\end{enumerate}

Note that the integrations over the reciprocal space patches in step 4 only involve the inversion of a matrix in angular momentum space
and therefore computational time is not significantly increased over the conventional KKR-CPA regardless of cluster size. This is in 
contrast to many other attempts such as the MCPA where the integration over the Brillouin zone requires the inversion of a super-matrix in 
cluster-site and angular momentum space for each value of $\mathbf{k}$. We also draw attention to the recent work of Maier and 
Jarrell~\cite{Ma:Ja:2q} where the DCA algorithm has been shown to converge more quickly than the MCPA algorithm, with corrections of order 
$1/L^2$ where $L$ is the linear size of the cluster.
 

\subsection{The cluster momenta $\{\mathbf{K}_n\}$}

\label{Kpoints}

Since the NLCPA maps the impurity cluster problem to the effective lattice problem in reciprocal space, it is important to realize that 
the real-space cluster must have periodic boundary conditions i.e. must preserve the translational symmetry of the lattice. Moreover, as
explained in Ref.~\onlinecite{Ja:Kr}, in order to obtain suitable reciprocal space patches centered at the cluster momenta 
$\{\mathbf{K}_n\}$, we must select the real-space cluster sites by surrounding them with a `tile' of size $L^D$ (where $D$ is the dimension)
which preserves the full point-group symmetry of the lattice and only clusters which satisfy this requirement are allowed.

The method for finding the corresponding cluster momenta $\{\mathbf{K}_n\}$ satisfying Eq.~(10) for a simple two-dimensional square lattice
has been described in Ref.~\onlinecite{Ja:Kr}. Here we generalize this method to the case of three-dimensional simple cubic (sc), 
body-centered cubic (bcc), and face-centered cubic (fcc) lattices commonly found in real disordered alloys. For the trivial case of $N_c=1$,
the real-space `tiles' with the required symmetry would simply be Wigner-Seitz cells surrounding each lattice point. For larger cluster sizes
we may take the `tiles' to be simple cubes of volume $L^3$ for each of these lattices. The smallest possible cluster sizes are given by 
considering $L=a$ (where $a$ is the lattice constant) and this yields $N_c=1$ for sc (trivial), $N_c=2$ for bcc, and $N_c=4$ for fcc 
lattices. The next set of allowed cluster sizes is given by considering $L=2a$ and this yields $N_c=8$, $N_c=16$, and $N_c=32$ for sc, bcc 
and fcc lattices respectively. 

The next step is to consider the `principal tiling vectors' $\{\mathbf{a}_1,\mathbf{a}_2,\mathbf{a}_3\}$ where 
$l\mathbf{a}_1+m\mathbf{a}_2+n\mathbf{a}_3$ for integers $l,m,n$ map out the centers of the real-space `tiles'. Applying the usual 
reciprocal space transformations of the form 
$\mathbf{b}_1= 2\pi\left(\mathbf{a}_2\times\mathbf{a}_3\right)\div\left(\mathbf{a}_1\cdot(\mathbf{a}_2\times\mathbf{a}_3)\right)$ etc.
gives us principal `coarse-grained' reciprocal space vectors $\{\mathbf{b}_1,\mathbf{b}_2,\mathbf{b}_3\}$. We take $N_c$ non-equivalent 
vectors (i.e. do not differ by a reciprocal lattice vector) of the form $l\mathbf{b}_1+m\mathbf{b}_2+n\mathbf{b}_3$ for integers $l,m,n$ 
to be our set of cluster momenta $\{\mathbf{K}_n\}$ and these will satisfy Eq.~(10). The reciprocal space patches surrounding these cluster
momenta will be simple cubes of equal volume $(2\pi/L)^3$ which together will fill out a volume the size of the first Brillouin zone i.e. 
$(2\pi/a)^3$ for sc, $2(2\pi/a)^3$ for bcc, and $4(2\pi/a)^3$ for fcc lattices respectively. Integrating over these patches is equivalent to
integrating over the first Brillouin zone of the lattice since $\mathbf{K}_n$ values and parts of patches lying outside of the first 
Brillouin zone can be translated through reciprocal lattice vectors (of the sc, bcc, or fcc lattice as appropriate) to lie within the first
Brillouin zone. In Tab.~\ref{table} we give some examples of sets of $\mathbf{R}_I$ and corresponding $\mathbf{K}_n$ values obtained using 
the above method.

\begin{table*}
\caption{Examples of sets of $\mathbf{R}_I$ and $\mathbf{K}_n$ values for lattices of the sc, bcc and fcc type}
\label{table}
\begin{tabular}{c|c|c|c|c} \hline\hline
 
lattice & $N_c$ & cube edge & $\mathbf{R}_I$ values (units of lattice constant $a$) & $\mathbf{K}_n$ values (units of $\frac{\pi}{a}$) \\
  
\hline\hline &&&& \\
 
sc  & 1  & $L=a$  & $(0,0,0)$ & $(0,0,0)$ \\  &&&& \\
 
\cline{2-5} &&&& \\

    & 8  & $L=2a$ & $\{\mathbf{R}_I^{sc}\}=$ $\{$ $(0,0,0)$, $(0,0,1)$, & $\{\mathbf{K}_n^{sc}\}=$ $\{$ $(0,0,0)$, $(0,0,1)$, \\ 
   
    &    &        & $(0,1,0)$, $(0,1,1)$, $(1,0,0)$, & $(0,1,0)$, $(0,1,1)$, $(1,0,0)$, \\
     
    &    &        & $(1,0,1)$, $(1,1,0)$, $(1,1,1)$ $\}$ & $(1,0,1)$, $(1,1,0)$, $(1,1,1)$ $\}$ \\

&&&& \\ \hline &&&& \\

bcc & 2  & $L=a$  & $(0,0,0)$, $(\frac{1}{2},\frac{1}{2},\frac{1}{2})$ & $(0,0,0)$, $(2,0,0)$ \\
 
&&&& \\ \cline{2-5} &&&&\\
    
    & 16 & $L=2a$ & $\{\mathbf{R}_I^{sc}\}$, $\{\mathbf{R}_I^{sc}+(\frac{1}{2},\frac{1}{2},\frac{1}{2})\}$ &
    					
    		    $\{\mathbf{K}_n^{sc}\}$, $\{\mathbf{K}_n^{sc}+(2,0,0)\}$ \\
   		     
&&&& \\ \hline &&&& \\	
    										  					
fcc & 4  & $L=a$  & $(0,0,0)$, $(\frac{1}{2},0,\frac{1}{2})$, $(\frac{1}{2},\frac{1}{2},0)$, $(0,\frac{1}{2},\frac{1}{2})$ &
 
				$(0,0,0)$, $(2,0,0)$, $(0,2,0)$, $(0,0,2)$ \\ 
				 
&&&& \\ \cline{2-5} &&&& \\
                                                                            
    & 32 & $L=2a$ & $\{\mathbf{R}_I^{sc}\}$, $\{\mathbf{R}_I^{sc}+(\frac{1}{2},0,\frac{1}{2})\}$, &
    		
		    $\{\mathbf{K}_n^{sc}\}$, $\{\mathbf{K}_n^{sc}+(2,0,0)\}$ \\	
    
    &    &        & $\{\mathbf{R}_I^{sc}+(\frac{1}{2},\frac{1}{2},0)\}$, $\{\mathbf{R}_I^{sc}+(0,\frac{1}{2},\frac{1}{2})\}$ &
    
    		    $\{\mathbf{K}_n^{sc}+(0,2,0)\}$, $\{\mathbf{K}_n^{sc}+(0,0,2)\}$ \\
           
&&&& \\ \hline\hline
    
\end{tabular}
\end{table*}


\subsection{Short-range order}

The principal advantage of the KKR-NLCPA over the conventional KKR-CPA is that it can be implemented for alloys in which short-range 
ordering or clustering is present. To deal with this situation one must include an appropriate weighting for each of the $2^{N_c}$ impurity 
cluster configurations in step 6 of the algorithm. Note that such short-range order will not destroy the translational invariance because it
will be restored by the configurational averaging. However when using some method to weight the configurations it is important to bear in 
mind the periodic boundary conditions imposed on the cluster as explained in the previous section. To illustrate the above feature of 
the theory we will describe short-range order with a one-dimensional example later in this paper.


\subsection{Calculating observables}

In order to calculate observables such as the configurationally-averaged densities and density of states, we need to know the 
configurationally-averaged Green's function calculated within the NLCPA. The expression for the Green's function before averaging is
given by
\begin{eqnarray}
	G(E,\mathbf{r}_i,\mathbf{r}_j')=\sum_{LL'}Z_L^i(E,\mathbf{r}_i)\tau_{LL'}^{ij}Z_{L'}^j(E,\mathbf{r}_j') \nonumber \\
	-\sum_LZ_L^i(E,\mathbf{r}_i)J_L^i(E,\mathbf{r}_i)\delta_{ij}
\end{eqnarray}	
where $L(=l,m)$ is an angular momentum index and $\mathbf{r}_i(\mathbf{r}_j')$ lies within the unit cell centered at site $i(j)$. 
$Z_L^i(E,\mathbf{r}_i)$ and $J_L^i(E,\mathbf{r}_i)$ are the regular and irregular solutions~\cite{Fau:St} respectively of the single-site 
problem at site $i$. In the next section we show calculations for the configurationally-averaged density of states for a one-dimensional 
model and so here we demonstrate explicitly how to take the configurational average of the site-diagonal Greens function. It should be
stressed that unlike in calculations based on tight-binding models, in the present multiple scattering theory we solve for and describe the
site to site fluctuations of the `orbitals' $Z_L^i(E,\mathbf{r}_i)$. In short we can calculate the density $\mathbf{r}$ point by 
$\mathbf{r}$ point. It is this feature of the present theory which requires the following careful averaging procedure.

The first step is to consider $\mathbf{r}$ and $\mathbf{r}'$ in the neighborhood of a cluster site $I$. Denoting a configuration of the 
remaining cluster sites by $\gamma$, as a generalization of Ref.~\onlinecite{Fau:St} we first average over the subset of possible lattice 
structures that leave the potential in site $I$ fixed:
\begin{widetext}
\begin{eqnarray}
	\langle G(E,\mathbf{r},\mathbf{r}')\rangle_I=\sum_{LL'}Z_L^I(E,\mathbf{r}_I)\left(\sum_{\gamma}P(\gamma | I)
	\langle\tau_{LL'}^{II}\rangle_{I,\gamma}\right)Z_{L'}^I(E,\mathbf{r}'_I) \nonumber \\
	-\sum_LZ_L^I(E,\mathbf{r}_I)J_L^I(E,\mathbf{r}'_I) 
\end{eqnarray}
\end{widetext}
where \,$\langle\tau_{LL'}^{II}\rangle_{I,\gamma}$\, is the path operator for paths starting and ending at site $I$ conditionally averaged so 
that the potential on site $I$ is known (to either be of type $A$ or $B$) and the configuration of the remaining sites in the cluster is 
known to be \,$\gamma$\,. $P(\gamma | I)$ is the probability that the configuration \,$\gamma$\, of the remaining cluster sites occurs given
the type of potential at site $I$. The final step is to average over the possible occupants of site $I$ itself:
\begin{widetext}
\begin{eqnarray}
	\langle G(E,\mathbf{r},\mathbf{r}')\rangle_I=\sum_{LL'} \left[P(A)Z_L^A(E,\mathbf{r}_I) \left(\sum_{\gamma}P(\gamma|A)
	\langle\tau_{LL'}^{II}\rangle_{A,\gamma}\right) Z_{L'}^A(E,\mathbf{r}'_I) \right.\nonumber \\
	 \left.+P(B)Z_L^B(E,\mathbf{r}_I)\left(\sum_{\gamma}P(\gamma |B)\langle\tau_{LL'}^{II}\rangle_{B,\gamma}\right) 
	 Z_{L'}^B(E,\mathbf{r}'_I) \right] \nonumber \\
	 -\sum_L\left[P(A)Z_L^A(E,\mathbf{r}_I)J_L^A(E,\mathbf{r}'_I)+P(B)Z_L^B(E,\mathbf{r}_I)J_L^B(E,\mathbf{r}'_I)\right] 
\end{eqnarray}
\end{widetext}
where $P(A)$ and $P(B)$ are the probablities that site $I$ is an $A$ atom or $B$ atom respectively (i.e. the concentrations of $A$ and $B$
atoms in the material). This can be rewritten as
\begin{widetext}
\begin{eqnarray}
	\langle G(E,\mathbf{r},\mathbf{r}')\rangle_I=\sum_{LL'} \left[\sum_{\gamma}P(A,\gamma)
	Z_L^A(E,\mathbf{r}_I)\langle\tau_{LL'}^{II}\rangle_{A,\gamma}\, Z_{L'}^A(E,\mathbf{r}'_I) \right.\nonumber \\
	\left.+\sum_{\gamma}P(B,\gamma)Z_L^B(E,\mathbf{r}_I)\langle\tau_{LL'}^{II}\rangle_{B,\gamma}\,
	Z_{L'}^B(E,\mathbf{r}'_I)\right] \nonumber \\
	 -\sum_L\left[P(A)Z_L^A(E,\mathbf{r}_I)J_L^A(E,\mathbf{r}'_I)+P(B)Z_L^B(E,\mathbf{r}_I)J_L^B(E,\mathbf{r}'_I)\right] 
\end{eqnarray}
\end{widetext}
where $(A,\gamma)$ and $(B,\gamma)$ denote cluster configurations with an $A$ atom and $B$ atom on site $I$ respectively. The above 
expression is still exact at this stage, however it can be simplified using the NLCPA. In this approximation, 
\,$\langle\tau_{LL'}^{II}\rangle_{I,\gamma}$\, is constructed using an `impurity' cluster of configuration $(I,\gamma)$ embedded in the 
NLCPA effective medium. By using Eq.~(4) to eliminate the cluster renormalised interactor from Eq.~(6), this is given by
\begin{equation}
	\langle\tau_{LL'}^{II}\rangle_{I,\gamma}=\left[\left(\widehat{\underline{\underline{\tau}}}^{-1}+
	\underline{\underline{t}}^{-1}_{cl,imp}-\widehat{\underline{\underline{t}}}^{-1}_{cl}\right)
	^{-1}\right]_{LL'}^{II}
\end{equation}
where the impurity cluster t-matrix \,$\underline{\underline{t}}_{cl,imp}$\, has configuration $(I,\gamma)$ and the notation implies taking 
the $I,I^{th}$ site and $L,L'\,^{th}$ angular momentum element of the super-matrix on the right hand side.

It does not matter which cluster site is chosen to be site $I$ in all the above formulae as $\widehat{G}(E,\mathbf{r},\mathbf{r}')$,
the resulting approximation to $\langle G(E,\mathbf{r},\mathbf{r}')\rangle$, is a translationally-invariant quantity. The density of 
states per site is given by
\begin{equation}
	\rho(E)=\frac{-1}{\pi}Im\int_{\Omega_I}\widehat{G}(E,\mathbf{r},\mathbf{r})\mathbf{dr}
\end{equation}
where the integral is over $\Omega_I$, the volume of site $I$.


\section{Results for one-dimensional model}

\label{results}

\begin{figure}
 \scalebox{0.35}{\includegraphics{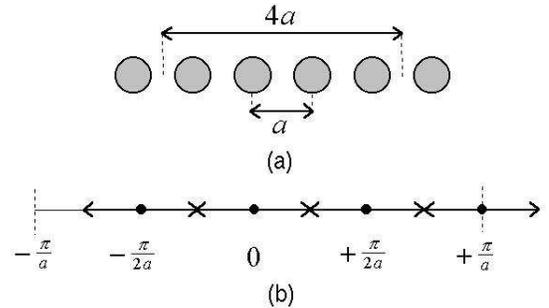}}
 \caption{A one-dimensional `tile' for a one-dimensional four effective site $(N_c=4)$ cluster is shown in (a). Its length $L$ is $4a$ 
 where $a$ is the lattice constant. The corresponding set of cluster momenta (denoted by dots) and reciprocal-space patches of length 
 $2\pi/L$ in relation to the first Brillouin zone are shown in (b).}
 \label{fig1}
\end{figure} 

\begin{figure}
 \scalebox{0.7}{\includegraphics{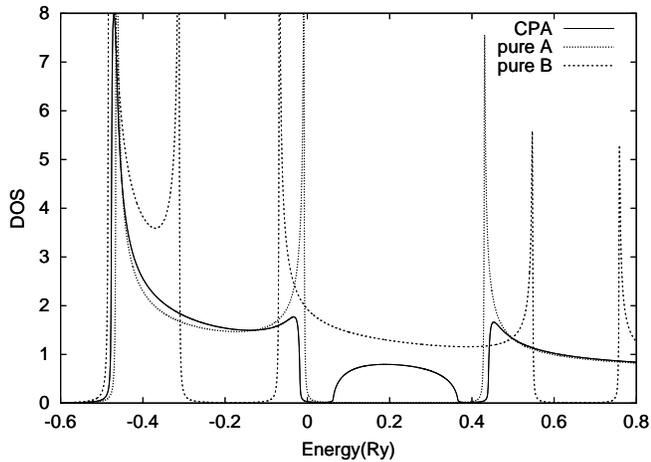}}
 \caption{Density of states for a one-dimensional model. Results shown are for a pure A lattice, a pure B lattice and a KKR-CPA 
 calculation for an $A_{80}B_{20}$ alloy.}
 \label{fig2}
\end{figure} 

\begin{figure}
 \scalebox{0.7}{\includegraphics{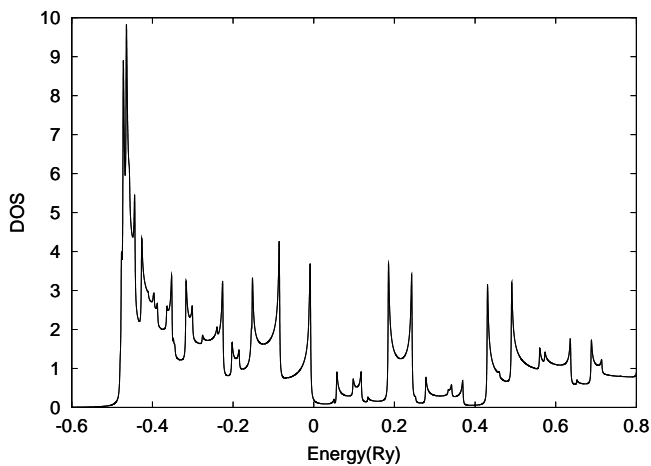}}
 \caption{A four site $(N_c=4)$ supercell calculation for the $A_{80}B_{20}$ alloy obtained by averaging over all $2^4$ possible
  configurations of an infinite periodic supercell containing four sites.}
 \label{fig3}
\end{figure} 

\begin{figure}
 \scalebox{0.7}{\includegraphics{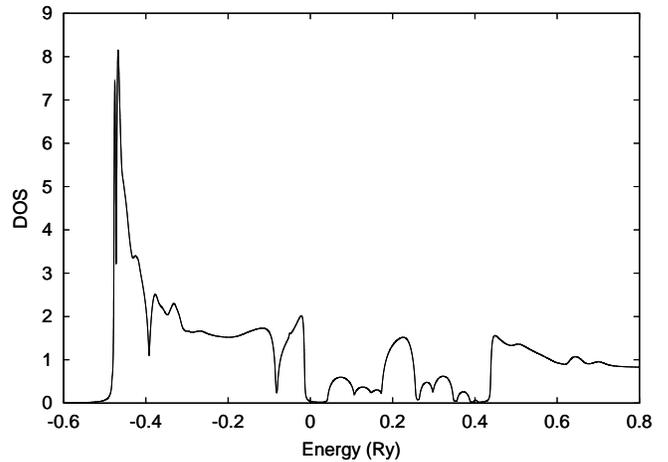}}
 \caption{A four site $(N_c=4)$ KKR-NLCPA calculation for the $A_{80}B_{20}$ alloy. Notice the improved structure and the partial
  filling in of the band gaps compared with the KKR-CPA calculation in Fig.~\ref{fig2}.}
 \label{fig4}
\end{figure}  

\begin{figure}
 \scalebox{0.7}{\includegraphics{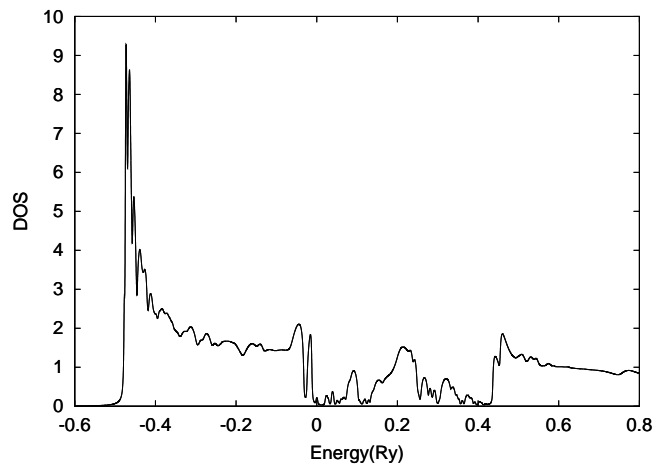}}
 \caption{An eight site $(N_c=8)$ KKR-NLCPA calculation for the $A_{80}B_{20}$ alloy. Notice the increasing density of states inside
  the CPA band gaps.}
 \label{fig5}
\end{figure}

\begin{figure}
 \scalebox{0.7}{\includegraphics{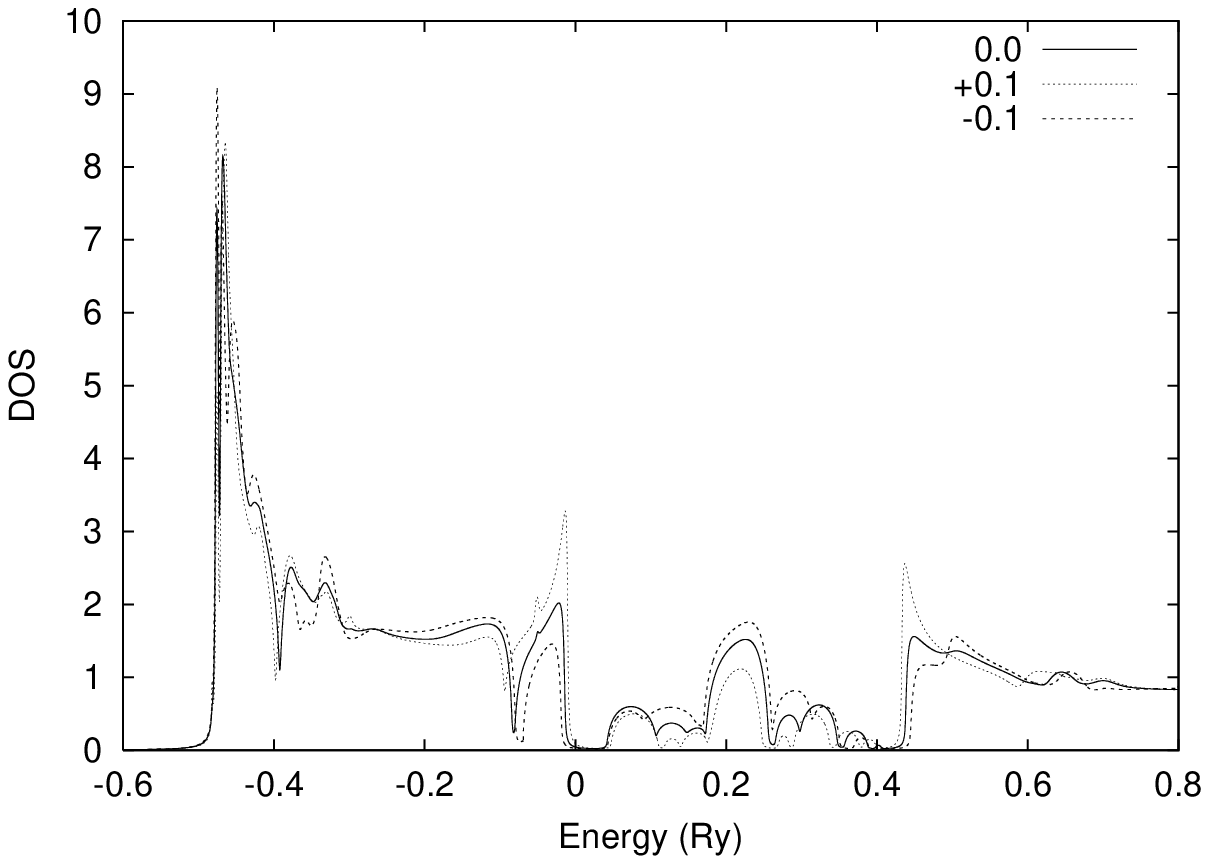}}
 \caption{Effects of short-range ordering $(\beta=-0.1)$ and clustering $(\beta=+0.1)$ on the four site $(N_c=4)$ KKR-NLCPA 
 calculation for the $A_{80}B_{20}$ alloy.}
 \label{fig6}
\end{figure} 

The multiple scattering theory in three dimensions envisions three-dimensional potential wells of finite range surrounding the atomic
nuclei, the famous muffin-tin potential. A very useful caricature of this realistic situation can be constructed in one dimension. Evidently
in this case the unit cell is a line segment and the potential wells are described by a function $V(x)$ of one spatial variable only.
Interestingly, one may regard the sign of $x$ as an angle and develop an analogue of the three-dimensional angular momentum expansion of the
usual KKR theory. Since it was first formulated by Butler~\cite{Butler} it has been made good use of by a number of authors, for example see
Refs.~\onlinecite{Sch:Gy:a,Sch:Gy:b,Sch:Gy:c}. While it is computationally simple, as one might expect, it is formally identical to KKR in 
three dimensions. For example there are two `angular momentum' values $L=0,1$ (and hence all NLCPA super-matrices in the algorithm have 
dimension $2\times N_c$) and there is an explicit expression for the structure constants. For a detailed description see 
Refs.~\onlinecite{Butler,Sch:Gy:a}. 

For a cluster of size $N_c$, the one-dimensional reciprocal space patches are simply defined by the points
\[ \mathbf{K}_n=\frac{(2n-N_c)\pi}{N_ca} \] where $a$ is the lattice constant and  $n=1,\ldots,N_c$. As an example, the real space 
`one-dimensional tiles' and reciprocal space patches for a four site $(N_c=4)$ cluster are shown in Fig.~\ref{fig1}.

We have carried out extensive numerical calculations of the configurationally-averaged density of states for a one-dimensional alloy over 
a wide range of parameters. In all cases we have found that the KKR-NLCPA systematically improves the density of states with increasing 
cluster size compared to the conventional KKR-CPA.

As a simple illustration we set the lattice constant to be $a=6.00$ a.u. and potential-well radius to be $r_{MT}=2.25$ a.u. for each of the
constituent potentials for $A$ and $B$ sites which are square potential wells of depth $-1.2$ Ry and $-0.6$ Ry respectively. The 
concentration of $A$ sites is taken to be $80\%$. The density of states for the electrons in lattices of purely A sites and purely B sites 
together with the CPA result for the $A_{80}B_{20}$ alloy is shown in Fig.~\ref{fig2}. In Fig.~\ref{fig3} we show for comparison the results
for a four site supercell calculation. This is obtained by considering an infinite periodic supercell containing four sites of a particular
configuration and then averaging over all $2^4$ possible configurations. The supercell calculation is not the exact result due to the finite
size of the supercell but it gives an indication of the type of structure to expect if we are to improve upon the CPA. A KKR-NLCPA 
calculation with a cluster size of four $(N_c=4)$ is shown in Fig.~\ref{fig4}. It is evident that much of the structure missing from the CPA
calculation which can be associated with energy bands of particular configurations of the supercell is reproduced here. States also appear 
in the band gaps either side of the impurity band centered at $0.2$ Ry which are absent in the CPA calculation. This is because the states 
near the band edges are the contributions of large clusters of like atoms and these cannot be dealt with by a single-site theory such as the
CPA. To investigate this further, an eight site $(N_c=8)$ calculation is shown in Fig.~\ref{fig5}. Clearly with increasing cluster size more
and more states enter the band gaps.

Next we illustrate the ability of the KKR-NLCPA to take into account the effects of short-range order. As mentioned earlier, the KKR-NLCPA 
can be implemented for arbitrary ensembles including those in which the occupancy of a site by an $A$ or $B$ atom is correlated to that of 
neighbouring sites. As a simple example we increase the probablity of an atom occupying a site by a factor $\beta$ if it follows a like atom
and decrease its probability by the same factor if it follows an unlike atom in the one-dimensional cluster. Thus positive values of $\beta$ 
correspond to short-range clustering and negative values of $\beta$ correspond to short-range ordering. As an example, for a four site 
impurity cluster of configuration $ABBA$ we have 
$P(ABBA)=\left(P(A)+\beta\right)\left(P(B)-\beta\right)\left(P(B)+\beta\right)\left(P(A)-\beta\right)$ where we have made use of the 
periodic boundary conditions imposed on the cluster. In Fig.~\ref{fig6} we show a four site $(N_c=4)$ KKR-NLCPA calculation for the density 
of states using the same parameters as before along with short-range order parameter values $\beta=-0.1$, $\beta=0.0$ and $\beta=+0.1$. 
Peaks which increase or decrease can be identified with specific cluster configurations and the increases or decreases in the amplitude of 
the peaks are consistent with the increased or decreased cluster probabilities.


\section{Conclusions}

We have presented the formalism for the KKR-NLCPA method which systematically improves upon the conventional KKR-CPA for describing
disordered systems on the basis of a first-principles description of the crystal potential. We have demonstrated its use on a one-dimensional
model and illustrated in detail the necessary coarse-graining procedure for real three-dimensional lattices. We have also shown how to 
calculate observable quantities with a view to combining the KKR-NLCPA with density functional theory. In the explicit calculations the 
emphasis was on the improved structure in the density of states with increasing cluster size due to non-local correlations, and a simple 
example of the ability of the KKR-NLPCA to model the effects of compositional short-range order.

In order for disordered systems to receive a `first-principles' description, electronic density functional theory~\cite{Dre:Gro} (DFT) needs 
to be combined with treatments of disorder. The self-consistent-field Korringa-Kohn-Rostocker coherent potential approximation 
(SCF-KKR-CPA)~\cite{St:Te:Gy,Jo:Ni:Pi:Gy:St} approach is a `first-pass' way at doing this and has been applied to a wide range of disordered 
alloys.~\cite{St:Jo:Pi,Abrikosov} It has also been adapted for the problem of itinerant magnets at finite temperatures whose `disordered 
local moment' spin fluctuations are handled using the CPA.~\cite{St:Gy,Ling} In principle a DFT calculation of a disordered system should
mean that separate SCF calculations are carried out to minimise the total energy for each disorder configuration individually and then an 
average taken over all disorder configurations. This is, of course, intractable and the strategy behind SCF-KKR-CPA calculations has been to 
minimise a functional for the averaged energy in terms of partially averaged charge (and spin) densities, i.e. the average of charge
densities arising from all configurations which have either an A or a B atom on one site. From this approach, however, it is not 
straightforward to include local environment effects such as electrostatic, `charge-correlation'~\cite{Jo:Pi,Ma:Wei:Zu,Abrikosov} Madelung 
and local lattice displacement effects in alloys or indeed short-range order effects in general. For disordered alloys, the implementation 
requires the solution of the Kohn-Sham equations for electrons moving in the disordered arrays of the effective potentials associated with A 
and B sites which are insensitive to their environments. In the context of finite temperature magnetism it means that the thermally induced 
spin fluctuations must be characterised as `local moments' associated with single sites and not larger clusters. In this paper we have taken 
the first steps in showing how these serious omissions can be rectified by presenting a scheme within the KKR framework which goes 
systematically beyond the CPA. 


\appendix

\section{}

Consider a finite-sized cluster $C$ of sites in the effective medium. Eq.~(2) for sites $i,j$ belonging to the cluster can be written in 
the form
\begin{equation}
	\widehat{\underline{\tau}}\,^{ij}=\widehat{\underline{t}}\,\delta_{ij}+\sum_{k\in C}\widehat{\underline{t}}\,
	\widehat{\underline{G}}(\mathbf{R}^{ik})\widehat{\underline{\tau}}\,^{kj}                        
	 +\sum_{k\not{\in}C}\widehat{\underline{t}}\,\widehat{\underline{G}}(\mathbf{R}^{ik})\widehat{\underline{\tau}}\,^{kj}
\end{equation}
where the sum over all sites $k$ has been split into those involving sites $k$ within the cluster and sites $k$ outside of the cluster.
It can be shown~\cite{Gonis,Gonis:ECM} that
\begin{equation}
	\sum_{k\not{\in}C}\widehat{\underline{G}}(\mathbf{R}^{ik})\widehat{\underline{\tau}}\,^{kj}=
	\sum_{l\in C}\widehat{\underline{\Delta}}\,^{il}\widehat{\underline{\tau}}\,^{lj}
\end{equation}
with the effective cluster renormalised interactor $\widehat{\underline{\Delta}}\,^{ij}$ given by the locator expansion
\begin{eqnarray}
	\widehat{\underline{\Delta}}\,^{ij}=\sum_{k\not{\in}C}\widehat{\underline{G}}(\mathbf{R}^{ik})\,\widehat{\underline{t}}\,
	\widehat{\underline{G}}(\mathbf{R}^{kj}) \nonumber \\
	+\sum_{k\not{\in}C,l\not{\in}C}\widehat{\underline{G}}(\mathbf{R}^{ik})\,\widehat{\underline{t}}\,
	\widehat{\underline{G}}(\mathbf{R}^{kl})\,\widehat{\underline{t}}\,\widehat{\underline{G}}(\mathbf{R}^{lj})+\cdots
\end{eqnarray}
Inserting (A2) into (A1) and using the notation that cluster sites are denoted by capital letters gives
\begin{equation}
	\widehat{\underline{\tau}}\,^{IJ}=\widehat{\underline{t}}\,\delta_{IJ}+\sum_{K}\widehat{\underline{t}}\,
	\left(\widehat{\underline{G}}(\mathbf{R}^{IK})+\widehat{\underline{\Delta}}\,^{IK}\right)\widehat{\underline{\tau}}\,^{KJ}
\end{equation}
which can be re-arranged as
\begin{equation}
	\widehat{\underline{\tau}}\,^{IJ}=\widehat{\underline{t}}_{cl}\,^{IJ}+\sum_{K,L}\widehat{\underline{t}}_{cl}\,^{IK} 
	\,\widehat{\underline{\Delta}}\,^{KL}\,\widehat{\underline{\tau}}\,^{LJ}
\end{equation}
to include the effective cluster t-matrix given by Eq.~(5).


\section{}

Firstly we note that the correlations between the cluster sites which we are aiming to reproduce in the effective lattice are of a finite 
range, say $\lesssim L/2$, where $L$ is the linear size of the `tile' surrounding the cluster sites (see Sec.~\ref{Kpoints}). The next step 
is to divide or `coarse-grain' the first Brillouin zone of the lattice of $N$ sites into $N_c=L^D$ patches of size $(2\pi /L)^D$ (where $D$
is the dimension) centered at the cluster momenta $\{\mathbf{K}_n\}$. 
We now consider each coarse-grained value \,$\widehat{\underline{\alpha}}(\mathbf{K}_n)$\, to be the average of 
\,$\widehat{\underline{\alpha}}(\mathbf{k})$\, over the $N/N_c$ lattice momenta \,$\tilde{\mathbf{k}}$\, within the patch surrounding the 
point $\mathbf{K}_n$:
\begin{equation}
	\widehat{\underline{\alpha}}(\mathbf{K}_n)=
	\frac{N_c}{N}\sum_{\tilde{\mathbf{k}}}\widehat{\underline{\alpha}}(\mathbf{K}_n+\tilde{\mathbf{k}})
\end{equation}

The fundamental assumption we make is to take these coarse-grained values \,$\widehat{\underline{\alpha}}(\mathbf{K}_n)$\, to be a 
good approximation to \,$\widehat{\underline{\alpha}}(\mathbf{k})$\, for the effective lattice.~\cite{Ja:DCA} This is because 
according to Nyquist's sampling theorem~\cite{Rao} in order to reproduce correlations of a finite range $(L/2)$ in real space, we only need 
to sample the first Brillouin zone of the lattice at intervals of \,$2\pi /L$\, i.e. at the cluster momenta $\{\mathbf{K}_n\}$. The real 
space correlation terms \,$\widehat{\underline{\alpha}}^{ij}(1-\delta_{ij})$\, are cut off if the distance between $i$ and $j$ is outside
the range of the cluster size. As $N_c\rightarrow\infty$ correlations over all length scales are treated since the maximum 
correlation length is proportional to the cluster size $N_c$.


\begin{acknowledgments}
	The authors would like to thank M.Jarrell for useful discussions. This work has been supported in part by EPSRC (UK).
\end{acknowledgments}




\end{document}